\documentclass[11pt]{article}
\usepackage{moriond,epsfig,amsfonts,latexsym}

\def\IZ{\mathbb{Z}}
\def\IR{\mathbb{R}}

\def\be{\begin{equation}}
\def\ee{\end{equation}}
\def\bea{\begin{eqnarray}}
\def\eea{\end{eqnarray}}

\begin{document}
\vspace*{4cm}
\title{PROSPECTS FROM STRINGS AND BRANES}

\author{ A.SEVRIN }

\address{Theoretische Natuurkunde, Vrije Universiteit Brussel, and
The International Solvay Institutes,\\
Pleinlaan 2, B-1050 Brussels, Belgium}

\maketitle\abstracts{
A brief, non-technical and non-exhaustive review of D(irichlet)-branes and (some) of their 
applications is given.}

\section{Introduction}
In this paper I will give a brief account of string theory with particular emphasis on D-branes and
their applications. A more extensive account can be found in a paper by Augusto Sagnotti and the
author which also provides a more comprehensive list of references.\cite{Sagnotti:2002yc} 
Because of the very nature of this
contribution, I will mainly cite review papers or books.\cite{Books on Strings} 
Further references to the original literature can be found in there.

\begin{figure}
\begin{center}
\psfig{figure=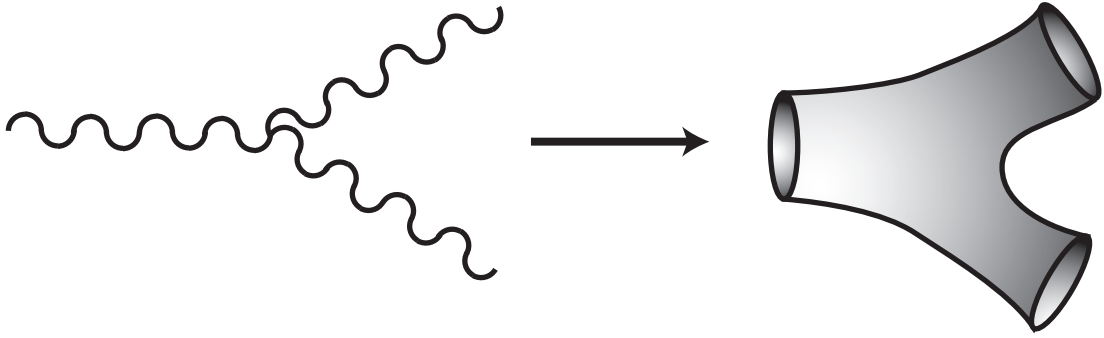}
\caption{By replacing point
particles with strings, a graviton three-point 
interaction e.g (left) is ``smeared''
such that in the resulting three-string vertex (right)
no localized interaction point can be found anymore.
\label{fig:graviton}}
\end{center}
\end{figure}

Einstein's general relativity gives a remarkably good {\em classical} description of the gravitational
interaction. However, any naive attempt to quantize general relativity fails as the theory is {\em 
non-renormalizable}. This means that the elimination of ultra-violet divergences -- endemic to quantum 
field theories -- necessitates the introduction of an infinite number of parameters, all to be
determined by experiments. This is clearly not an acceptable situation. Modifying the ultra-violet 
structure 
of the theory by smearing out the interactions cures this problem. In casu, we replace point particles 
by tiny strings as shown on figure \ref{fig:graviton}.

String theory is not new. In fact it finds its roots in the late sixties, early seventies, as 
an attempt to describe the strong interaction. However, in that
context it encountered a grave problem: in the particle spectrum of a string theory one always finds
a massless spin 2 particle. With the advent of QCD, string theory was abandoned as a theory of the
strong force but it lived on as a candidate for a theory of quantum gravity (the graviton {\em is} a 
massless spin 2 particle). 

The supersymmetric version of string theory is ultra-violet finite. It contains both the gravitational 
and the other interactions in such a way that in the infra-red regime they effectively reduce to (a 
supersymmetric extension of) general relativity and gauge theories. While 
this looks very encouraging, one has to
face the problem that consistency of string theory at the quantum level requires it to live
in 9 space-like 
dimensions and 1 time-like dimension. In 
other words: the theory only exists in 10 (or $9+1$) dimensions. 
This obvious problem can be solved in two ways (or a 
combination of these):
\begin{itemize}
\item Make the superfluous dimensions very small. This is known as {\em compactification}. Compare 
this to a garden hose which, from far away, looks like a one-dimensional object. The precise shape and 
volume of the compact dimensions determines to a large extent the physics in the four uncompactified 
dimensions. This immediately leads to another problem. While string theory puts very severe 
consistency requirements on the compact space, numerous solutions are known. A 
physical principle to select the ``right one'' is still lacking.
\item Make the extra dimensions very dark. This covers to a large extent the {\em brane-world} 
scenarios. The main idea is that the gauge interactions are confined to our four-dimensional
world while the gravitational interaction propagates in the full ten-dimensional space. This option is 
reviewed in more detail in the contribution of Lisa Randall in this volume.
\end{itemize}
As mentioned, string theory allows for a very large number of different ways to realize either one or 
a combination of those options. The quest for a selection mechanism requires insight 
in the non-perturbative properties of 
string theory. This is a highly non-trivial task as a string theory is essentially defined as a set
of self-consistent Feynman rules and as consequence it is purely perturbative. With the discovery of 
D-branes in 1995, certain non-perturbative issues became addressable.

\section{D-branes}
Strings occur in two versions: closed and open strings. Roughly speaking, one has that closed
strings carry the gravitational interaction and the open strings carry the gauge interactions.
While closed strings can freely propagate in space, the modern point of view is that the end points of
open strings are ``stuck'' on $p$-dimensional hypersurfaces, where $p\in\{1,2,\cdots,9\}$. These 
hypersurfaces are known as Dp-branes. They are dynamical but they are extremely heavy in the 
perturbative regime of string theory (their tension or energy per unit of volume is inversely 
proportional to the string coupling 
constant): they are {\em solitons}. A D0-brane is a point-like object, a D1-brane a string-like 
object, a D2-brane a membrane, ... Just as a propagating point particle sweeps out 
a curve -- the world-line -- in
space-time, a Dp-brane sweeps out a $p+1$-dimensional 
volume -- the world-volume -- in the 10-dimensional 
space-time. The effective dynamics on the world-volume is then described by a $p+1$-dimensional field
theory.

What are the degrees of freedom in this field theory? 
Looking at a single Dp-brane, one finds that 
the bosonic degrees of freedom are a $U(1)$ gauge field (a photon) and $9-p$ scalar fields together
with their fermionic partners. Both types of 
fields arise from the fluctuations of the open strings ending on 
the brane. The photon corresponds to fluctuations longitudinal to the brane while the scalar
fields are associated to the fluctuations of the string transversal to the brane. In the infra-red, 
the effective field theory on the brane is simply a supersymmetric version of Maxwell theory coupled 
to $9-p$ scalar fields.

\begin{figure}
\begin{center}
\psfig{figure=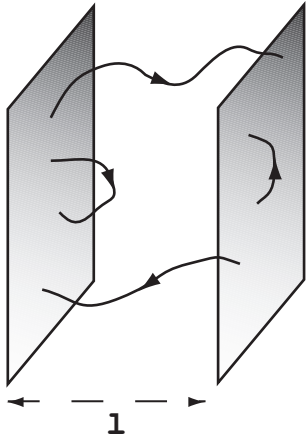}
\caption{Open (oriented) strings can attach themselves in four different ways to two parallel 
D-branes. The mass of the gauge field associated to each of these open 
strings is proportional to the shortest distance between the branes they connect. Whenever $l\neq 0$,
one has two massless gauge fields, with a corresponding unbroken $U(1)\times U(1)$
symmetry, and two additional massive $W$-like fields. On the other hand, when $l\rightarrow 0$ the 
two $W$-like fields become massless as well,
and the gauge symmetry is enhanced to $U(2)$. This yields a geometric 
setting for the Brout-Englert-Higgs mechanism. The Higgs scalar
describes the fluctuations of the branes relative to one another with its
vacuum expectation value corresponding to the relative distance, $l$, between the branes. 
\label{fig:2branes}}
\end{center}
\end{figure}

Once more D-branes are present, the situation becomes interesting. The mass of an open string is 
proportional to the shortest distance between the two branes it connects. As explained in figure 
\ref{fig:2branes}, when several, say $n$ branes coincide, additional massles gauge fields appear and
the gauge symmetry grows from $ (U(1))^n$ to $U(n)$.

\section{Applications}
\subsection{(Non-)abelian gauge theories and their solutions}
The effective field theory describing $n$ D-branes is in leading order a supersymmetric 
gauge theory with \footnote{Using somewhat more intricate constructions, other gauge groups are 
possible as well.} gauge group $U(n)$. In this way, D-branes provide an excellent laboratory to
study various aspects and solutions of gauge theories in a very geometric setting. The
Brout-Englert-Higgs mechanism, discussed in the previous section, provided a first example.

\begin{figure}
\begin{center}
\psfig{figure=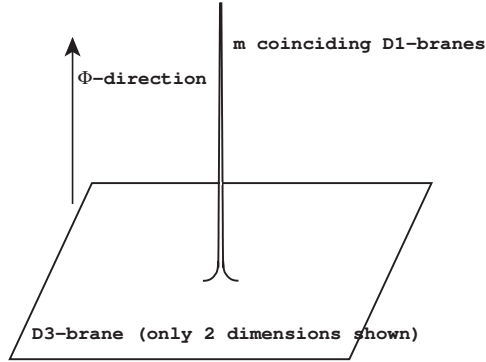}
\caption{The Dirac magnetic monopole with flux $m$ is realized in
string theory as a D3-brane with $m$ D1-branes perpendicular to it.
\label{fig:bion}}
\end{center}
\end{figure}

Another simple and instructive example is Dirac's magnetic monopole.\cite{gibbons} Consider (in 3 
dimensions) an electro-magnetic potential of the form,
\begin{eqnarray}
A_x^{(\pm)}=-\frac m 2 \frac{y}{r(z\pm r)},\quad
A_y^{(\pm)}=+\frac m 2 \frac{x}{r(z\pm r)},\quad
A_z^{(\pm)}=0,
\end{eqnarray}
where $r\equiv \sqrt{x^2+y^2+z^2}$ and $A^{(+)}$ ($A^{(-)}$ resp.) is defined 
for $ \theta > -\varepsilon $ ($ 
\theta < \varepsilon $ resp.), with $ \theta $
the azimuthal angle and $ \varepsilon >0$ and small. If one requires that on the overlap of the two 
patches ($ - \varepsilon < \theta < \varepsilon $), $A^{(+)}$ and $A^{(-)}$ are related by a single 
valued gauge transformation, one obtains the celebrated Dirac quantization condition: $m\in \IZ$.
The magnetic field $\vec B= \nabla\times\vec A$, satisfies,
\begin{eqnarray}
\nabla\cdot \vec B= 2 \pi m \,\delta (\vec x),
\end{eqnarray}
thus signalling the presence of a magnetic monopole at the origin.

In the language of D-branes, one realizes this as follows. One starts with a single D3-brane stretched 
in the $x-y-z$ directions, together with a scalar field $ \Phi $ which has the background value,
$ \Phi = m/2r$. The field equations are satisfied if the magnetic field is given by $ \vec B =
-\nabla \Phi $ which is precisely Dirac's monopole field. As explained before, a scalar field 
corresponds to a direction perpendicular to the D3-brane. A careful analysis of the various charges 
shows that this is a stable D-brane configuration consisting of a single D3-brane with $m$ D1-branes 
perpendicular to it thereby giving yet another interpretation/derivation of Dirac's quantization 
condition. The system is illustrated in figure \ref{fig:bion}.

In fact, to the authors knowledge, {\em almost all} monopole and instanton configurations find a 
natural 
realization in terms of D-branes. This setting provides not only a classification of such solutions
but elucidates many of their properties (e.g. the ADHM construction of instanton solutions) as well. 
In fact, only one class of gauge configurations
escaped a D-brane interpretation: the octonionic monopoles in 7 and the octonionic instantons in 8 
dimensions. However, even these solutions might very well find their place in a D-brane 
context.\cite{prep}

\subsection{Black Holes}
A spectacular application of strings and branes occurs in the study of black holes.\cite{BH}
A static, isotropic object of mass $M$ with a radius $R$, smaller than the Schwarzschild radius 
$R_s=2G_NMc^{-2}$ ($G_N$ is Newton's constant) is a (Schwarzschild) black hole. The (imaginary) sphere 
with radius $R_S$ 
around the black hole is called the event
horizon. Physics inside the event horizon is completely disconnected from physics outside it. 
Taking quantum mechanics into account, Hawking showed that black holes are not really black
but that they radiate thermally with temperature,
\begin{eqnarray}
T_H\ =\ \frac{c^3\, \hbar}{8\pi k_B G_N M}\ , \label{temp}
\end{eqnarray}
where $k_B$ denotes Boltzmann's constant. Using the second law of thermodynamics, one finds
the entropy $S_H$,
\begin{eqnarray}
\frac{1}{k_B}\; S_H\ =\ \frac{4\pi G_N M^2}{c\hbar}\ = \
\frac 1 4 A_H \, \ell_{Pl}^{-2} \ , \label{schwarz}
\end{eqnarray}
where $A_H$ is the area of the horizon in Planck units and
$\ell_{Pl}$ is the Planck length $\ell_{Pl}=\sqrt{G_N \hbar/c^3}\approx  
\ 10^{-33}\,cm$. The fact that the entropy is 1/4 of the surface of the horizon in Planck units
is a universal behavior of all black holes. Furthermore, as was shown by 't Hooft, a black hole is the 
physical system which maximizes the entropy in a given volume. These simple observations raise three 
profound questions.
\begin{enumerate}
\item The entropy of a black hole is characterized by a few macroscopic 
quantities. In our example there is only
the mass $M$, the more general case can have some charges, angular momentum, ... 
Since Boltzmann, we know that entropy measures the degeneracy of microstates 
in some underlying microscopic description of the system.
Since the entropy (\ref{schwarz})
of a black hole is an  unusually large number,
how can one realize such a wealth of microstates?
\item Any object coming from outside and 
crossing the horizon is trapped inside it forever, leaving only thermal radiation behind. 
A black hole is a very simple object: no matter how diversified the objects absorbed by it, the result 
is characterized by a few macroscopic quantities.
This seems
to imply that the S-matrix of a system containing a black 
hole seems not unitary anymore, thus violating one of the basic axioms of quantum
mechanics (the {\em information paradox}).
\item As a black hole maximizes the entropy within a given volume, it is highly unusual that the 
entropy is proportional to the surface of the horizon rather than to its volume. This led 't Hooft and 
Susskind to the holographic principle: any theory with gravity (and as a consequence with black holes) 
in a given volume should somehow be equivalent to a theory without gravity (hence without black holes) 
living on the boundary of the volume. While both attractive and spectacular, one would like to have
concrete realizations of the holographic principle.
\end{enumerate}
For a certain class of black holes, the extremal {}\footnote{Extremal black holes have zero 
temperature and thus can be viewed as ``elementary particles''. A 
particular example is e.g. an electrically charged black hole where the mass/charge ratio has been 
fine-tuned such that the gravitational collapse precisely compensates the electro-static repulsion.} 
and near-extremal black holes, the first two questions were solved. Starting from the effective field 
theory describing the infra-red regime of a certain string theory (a specific supersymmetric extension 
of general relativity) one looks for black hole solutions characterized by their energy and some 
charges. Performing Hawking's program yields then an expression for the entropy. Subsequently one 
constructs within the corresponding string theory stable D-brane configurations wrapped around the 
compact dimensions with open strings ending on them which give rise to the same energy and charges. 
Usually their are numerous configurations giving rise to the same macroscopic numbers and it is quite 
spectacular that the resulting degeneracy exactly reproduces the macroscopically calculated entropy. 

\begin{figure}
\begin{center}
\psfig{figure=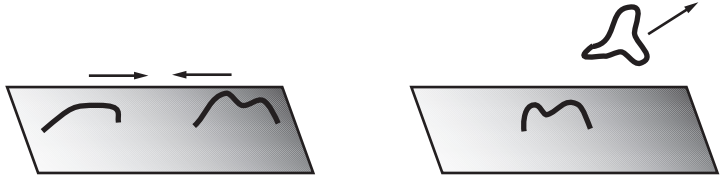}
\caption{The D-brane picture of Hawking radiation. A pair of open
strings collide, giving rise to a closed string that leaves the brane.
As a result, Hawking radiation reaches the bulk via the emission of
closed strings. 
\label{fig:bh}}
\end{center}
\end{figure}

For a near-extremal black hole the radiation of the system can be studied as well. Hawking radiation 
turns out, as shown in figure \ref{fig:bh}, to arise from the annihilation of open strings resulting 
in an 
open string remaining on the brane and a closed string leaving the brane. This radiation is exactly 
thermal with both temperature and radiation in perfect agreement with Hawking's calculation. By 
construction this approach is unitary and the apparently lost information appears to reside in the 
D-branes.
\subsection{Holography and the AdS/CFT correspondence}
The third problem cited in the previous section -- holography -- can as well be addressed in the 
context of D-branes.\cite{adscft} Maldacena considered a stack of $m$ D3-branes in IIB string theory 
in flat space. 
Taking the low energy limit in a specific way one finds that the bulk physics (containing a
non-interacting version of gravity) 
decouples from the brane physics (a $U(m)$ $N=4$ supersymmetric Yang-Mills theory in $3+1$ dimensions 
\footnote{$N=4$ denotes the number of supersymmetries, the theory has four times as much supersymmetry 
than e.g. the MSSM.}). Alternatively, looking at IIB supergravity (the effective field theory 
describing the infra-red regime of IIB superstrings) one constructs a solution having the same quantum 
numbers as the stack of D3-branes. The decoupling limit turns out in this case to be the near-horizon 
limit which has the topology of $ AdS_5\times S^5$, with $S^5$ the 5-dimensional sphere and $AdS_5$ 
5-dimensional anti-de Sitter space. For our purposes, it is sufficient to know that the latter is a 
manifold with a negative cosmological constant which has 4-dimensional Minkowski space as its 
boundary. The previous observations led Maldacena to the conjecture that string theory on $AdS_5\times 
S^5$ is {\em equivalent} or {\em dual} to $U(m)$ $N=4$ supersymmetric Yang-Mills theory living
on the boundary, i.e. $3+1$ dimensional Minkowski space. The concrete mapping between observables at 
both sides of this duality were found thereby providing a first concrete example of holography. While 
this is still a conjecture, it passed numerous tests and checks. The map between both theories is 
quite 
remarkable as the supergravity description of string theory (a purely classical limit) corresponds to 
the strongly coupled limit of the corresponding $U(m)$ gauge theory. In other words, classical 
calculations in supergravity yield non-perturbative information on a gauge theory. Since this seminal 
examples, many other instances of holography were found and the gravity/gauge map has become a 
powerful tool to study previously inaccessible features of gauge theories (e.g. Dijkgraaf and
Vafa mapped the full non-perturbative holomorphic sector of $N=1$ supersymmetric gauge theories to
a classical matrix model).

\subsection{Cosmology}
Observational evidence strongly points to the fact that the expansion of our universe
is presently in an accelerating phase.\cite{versnelling}
Let us first look at some elementary issues.\cite{cosmo} Approximating
the universe by a perfect fluid (characterized by an energy density $ \rho $ and a pressure $ p$), 
one finds that it is described by the Robertson-Walker line element,
\begin{eqnarray}
d \tau ^2= dt^2-R(t)^2\left(dr^2+r^2d \theta ^2+ r^2\sin^2 \theta d \phi ^2\right), 
\end{eqnarray}
where we assumed a flat topology for space (as is favored by the data). 
The scale factor $R(t)$ is determined by the Einstein equations which reduce to,
\begin{eqnarray}
&&\dot R^2= \frac{8 \pi G_N}{3}R^2 \left(\rho + \frac{ \lambda }{8\pi G_N}\right), \nonumber\\
&&\dot \rho +( \rho +p) \frac{3\dot R}{R}=0.\label{fl}
\end{eqnarray}
with $ \lambda $ the cosmological constant. In order to solve these equations, one needs an equation 
of state, i.e. a relation between the pressure $p$ and the energy density $ \rho $. The simplest 
ansatz is a constant linear relation,
\begin{eqnarray}
p(t) = \alpha \, \rho (t)&\Rightarrow& \rho = \rho _0 \frac{R_0^{3( \alpha +1)}}{R^{3( \alpha +1)}},
\label{eos}
\end{eqnarray}
with $ \alpha \in \IR$. The subindex $0$ denotes the present value of the corresponding quantities.
For non-relativistic dust one has $ \alpha =0$, radiation gives $ \alpha =1/3$ 
and a positive cosmological constant corresponds to $ \alpha =-1$. Using eq. (\ref{eos}) in eq.
(\ref{fl}) yields,
\begin{eqnarray}
\alpha \neq -1&&R(t)\propto t^{2/3( \alpha +1)}, \nonumber\\
\alpha =-1&& R(t)\propto e^{\sqrt{ \lambda }t/\sqrt{3}}.
\end{eqnarray}
Note that a positive cosmological constant ($ \alpha =-1$) gives rise to an exponentially growing 
expansion, also known as {\em inflation}.
From eq. (\ref{fl}) we immediately derive an expression for the change in the 
expansion rate,
\begin{eqnarray}
\frac{\ddot{R}}{R}=- \frac{4\pi G}{3}( 1+3\alpha)\rho + \frac{ \lambda }{3}.
\end{eqnarray}
One notices that a positive cosmological constant ($ \lambda >0$) results in acceleration. So does any 
``matter'' (meaning that $ \rho >0$)
which satisfies the equation of state, eq. (\ref{eos}) with $ \alpha <-1/3$. Present data favors 
$-1.62< \alpha < -0.74$. Models with $ \alpha <-1$ describe so-called phantom matter which will not 
further be discussed here \footnote{Phantom matter corresponds to e.g. scalar fields with the wrong 
sign for the kinetic energy. The reason why these scenario's are anyway studied is the possibility 
that the resulting instability might last much longer than the age of the universe.\cite{carroll} }.
If the acceleration is 
caused by a positive cosmological constant ($ \alpha =-1$), one has to face the cosmological constant 
problem. Indeed, the observed vacuum energy density associated to it is of the order $\rho_{vac}\sim  
\left(10^{-3}\,eV\right)^4 $. If one compares this to the Planck scale, $(10^{19} GeV )^4$, one finds 
an $ 
{\cal O}(10^{124})$ mismatch! Comparison with the supersymmetry breaking breaking scale, 
$(10^{3} GeV)^4$, improves the situation but still leaves an ${\cal O}(10^{60})$ discrepancy! In fact 
it turns out to 
be very hard to accommodate for a very small but non-zero cosmological constant. Because of this, the 
anthropic principle currently undergoes a revival...
 
\begin{figure}
\begin{center}
\psfig{figure=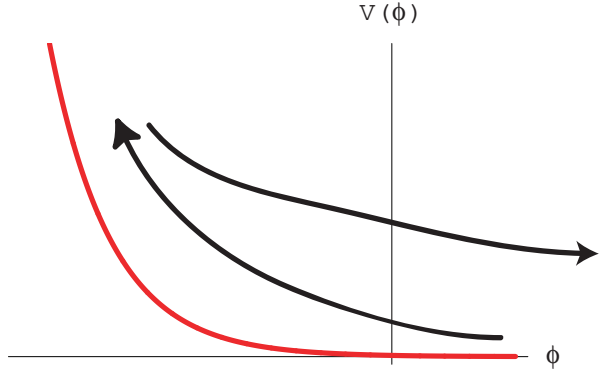}
\caption{A typical potential of the form $V\propto e^{-2a \phi }$. Initially the field starts from the 
right 
with a large negative velocity. It rolls up the hill, looses kinetic energy and enters a transitional 
accelerating epoch. During this period the constants of nature vary extremely slowly. 
\label{fig:potential}}
\end{center}
\end{figure} 
 
Turning to string theory, the situation even worsens. In 10- or 11-dimensional supergravity (the 
supersymmetric extensions of general relativity which describe the infra-red regime of string theory), 
the second derivative of the scale factor is directly related to the Ricci tensor, $\ddot R(t) = 
-R_{00}=-4\pi(T_{00}+g^{ij}T_{ij})$ with $T$ the energy-momentum tensor and $g$ the metric. Both 10- 
and 11-dimensional supergravity satisfy the {\em strong 
energy condition}, $R_{00}>0$, and as a consequence can never accommodate an accelerating universe.
Gibbons and later Nu\~nez and Maldacena studied 11- and 10-dimensional supergravity compactified to 
four dimensions on a static 7- or 6-dimensional space with metric,
\begin{eqnarray}
ds^2_{11,10}= \omega ^2(y) ds_4^2(x)+ds_{7,6}^2(y),
\end{eqnarray}
where $y$ collectively denote the compact coordinates. A straightforward calculation shows then that
$ R^{(11,10)}_{00}\geq 0\Rightarrow R_{00}^{(4)}\geq 0$. In other words, this looks as if string 
theory can never accommodate for an accelerating universe. There are two ways out. Ultra-violet 
effects 
might modify the behavior of the theory in such a way that acceleration does become possible. As the 
full ultra-violet behavior of string theory is not known yet, we leave this possibility for what it 
is. The other way is to reassess the premises of the no-go theorem. Doing so one immediately notices
the word {\em static}. 

Before delving deeper, let me mention that the shape, volume, ... of the compact space in string 
theory is steered by scalar fields (which generically are ubiquitous in string theory).
Consider the simple case of a single, uniform scalar field $ \phi $ with a potential $V( \phi )$, 
living in a Robertson-Walker background. Its equation of motion is given by,
\begin{eqnarray}
\ddot \phi =-V( \phi )'-3 H \dot\phi,\qquad H\equiv \dot R /R,\label{eoms}
\end{eqnarray}
where the dot denotes a time derivative and a prime a derivative with respect to $ \phi $.
The Robertson-Walker background provides a time-dependent friction term. The energy
density and pressure for this system are given by,
\begin{eqnarray}
\rho =\frac 1 2 \dot \phi {}^2+V( \phi ),\qquad p= \frac 1 2 \dot \phi {}^2-V( \phi ).
\end{eqnarray}
One notices that if both the kinetic energy is sufficiently small and the potential $V( \phi )$ is 
positive, acceleration occurs. The previously 
stated no-go theorem required a static compactification, in other words $ \phi $ is time independent. 
So from the equation of motion (\ref{eoms}) one sees that the no-go theorem implies that  
the potential has no stationary points with $V>0$. Allowing non-static compactifications yields
a way around the no-go theorem. 

A typical situation in string theory \cite{Townsend} -- arising both in flux and in hyperbolic 
compactifications -- is a potential $ V( \phi )= b\, e^{-2\,a\, \phi },$ with $ b>0$. For hyperbolic 
compactifications one has $ 1\,<\,a\,<\, \sqrt{3}$, while for flux compactifications $a>\sqrt{3}$ 
holds. At a far past, $\dot \phi \ll 0$ and $ \phi \gg 0$. As shown in figure \ref{fig:potential}, 
when time evolves, the field rolls up the hill while $\dot \phi $ approaches zero. Around the turning 
point we get a transient accelerating phase (during which the constants of nature, determined by the 
moduli fields are nearly constant). While this might be an explanation for the current acceleration, 
one could wonder whether this scenario might also explain the initial inflation in the early universe. 
A detailed investigation shows that it cannot as the number of e-foldings is far less than what is 
required to solve the horizon and flatness problems.
More subtle models which include orientifold planes and branes and which cannot so easily be captured 
in a supergravity language can actually reproduce intitial inflation.\cite{kklt} \cite{eva}
\subsection{Particle physics}
From the very beginning of string theory as a quantum theory of gravity, serious efforts were invested
in the study of its consequences for particle physics phenomenology. Before the advent of D-branes,
the efforts were concentrated on model building starting from the heterotic string string. Now that
we have D-branes, new possibilities open up, some of which are reviewed elsewhere in this volume.

We already mentioned the flux compactifications.\cite{eva} In these models the moduli are fixed by 
turning on 
fluxes for certain generalized gauge configurations (essentially for the so-called Ramond-Ramond 
fieldstrengths). This approach which has many interesting applications (see e.g. the previous 
subsection) has the drawback that the generation of chiral fermions is very hard. More involved model 
building along these lines is in full development. 

Another intriguing development are the so-called intersecting brane worlds.\cite{intersecting} 
In these, various stacks 
of branes are considered (the strong, the weak and the electro-magnetic) which have a 3-dimensional 
intersection - our world - where all three forces are simultaneously present. It turns out that it is 
easy to generate the Standard Model gauge group, 3 families of quarks and leptons, ... Unfortunately, 
there are many ways to achieve this. Generic features of these models are the presence of right-handed 
neutrinos and two or more Higgs scalars. An important difficulty is the issue of stability. (In order 
for this to work the branes have to intersect in a very specific way, however the branes tend to 
recombine.) Furthermore, the status of the hierarchy problem -- at least for toroidal and orbifold 
compactifications -- is unclear. Finally -- and related to the stability issue -- the construction of 
an intersecting brane world which yields the MSSM was till recently an open problem. 
However, very 
recently there was serious progress towards solving this problem.\cite{ott} 
At present a lot of effort is 
dedicated towards the construction of the low-energy effective action and the identification of the 
generic ``beyond-the-Standard-Model'' features of these models.

\section{Conclusions}
From the previous, it is clear that string theory accounts for several successes. Indeed, both the
microscopic understanding of (a class of) black holes and the realization of holography are
highlights. The close relation of D-branes with gauge theories provides a novel way of
studying various aspects of gauge theories from a geometric perspective. However, from a purely
particle physicists point of view, one has to admit that concrete qualitative, let alone quantitative
post- or predictions are not yet in sight. While both flux compactifications and intersecting brane
worlds are very valuable ideas which are thoroughly being explored, a mechanism for selecting the 
``right'' vacuum is not yet available. 

In the absence of this, one has recently started to explore an 
alternative way to arrive at predictions, which is the study of the so-called {\em string theory 
landscape}.\cite{landscape} 
In this statistical approach one counts the number of vacua having more or less the same physical 
properties, i.e. having the same values for certain fundamental parameters. (Because of e.g. hidden 
sectors, there might be a very large number of different vacua yielding all similar four dimensional 
physics.) One would expect that the most probable value for these parameters are those which are 
realized by the largest number of vacua.
While this approach is still in its infancy -- only relatively simple classes of models have already 
been 
investigated -- the first results are not really encouraging: neither low energy supersymmetry 
breaking 
nor large extra dimensions are favored. But then again, it will take another year or two of research
along these lines before hard statements can be made.

Finally, very recent ideas where the possibility of very large relic strings in the cosmos might lead 
to concrete predictions testable in gravitational wave detectors such as LIGO. \cite{cosmicstring}

\section*{Acknowledgments}
This work was supported in part by the ``FWO-Vlaanderen'' through 
project G.0034.02, in part by the Belgian Federal Science Policy Office 
through the Interuniversity Attraction Pole P5/27 and in part by the 
European Commission RTN programme HPRN-CT-2000-00131, in which the 
author is associated to the University of Leuven.

\end{document}